\documentclass[9pt,twocolumn,twoside]{opticajnl}
\journal{opticajournal} 

\setboolean{shortarticle}{true}

\usepackage{amsmath, amssymb, amsfonts}
\usepackage{siunitx}    
\usepackage{physics} 
\AtBeginDocument{\RenewCommandCopy\qty\SI}
\usepackage{graphicx}
\usepackage{microtype}
\usepackage{booktabs}
\usepackage{xcolor}
\usepackage[capitalise]{cleveref}
\usepackage{enumitem}
\setlist{nosep}

\usepackage{acro}
\usepackage{subcaption}
\usepackage{hyperref}

\usepackage{lineno}

\DeclareAcronym{OPO}{short=OPO, long=optical parametric oscillator}
\DeclareAcronym{SHG}{short=SHG, long=second-harmonic generation}
\DeclareAcronym{PPKTP}{short=PPKTP, long=periodically poled potassium titanyl phosphate}
\DeclareAcronym{QPM}{short=QPM, long=quasi-phase matching}

\DeclareAcronym{AR}{short=AR, long=anti-reflection}
\DeclareAcronym{HR}{short=HR, long=high-reflection}
\DeclareAcronym{PR}{short=PR, long=partial-reflection}

\DeclareAcronym{SPDC}{short=SPDC, long=spontaneous parametric down-conversion}
\DeclareAcronym{CV}{short=CV, long=continuous-variable}
\DeclareAcronym{GKP}{short=GKP, long=Gottesman--Kitaev--Preskill}

\title{Phase-Sensitive Crystal-Edge Effects in Linear Optical Parametric Oscillators: Why Nominally Identical Squeezers Behave Differently}

\author[1,2,*]{Jonas Junker}
\author[1,$\dagger$]{Donghwa Lee}
\author[1]{Reda Louahaj}
\author[1]{Oscar Cordero}
\author[1]{Romain Brunel}
\author[1]{Jens A. H. Nielsen}
\author[1]{Ulrik L. Andersen}
\author[1,$\ddagger$]{Jonas S. Neergaard-Nielsen}

\affil[1]{Center for Macroscopic Quantum States (bigQ), Department of Physics, Technical University of Denmark, Fysikvej, 2800 Kongens Lyngby, Denmark}
\affil[2]{Friedrich Schiller University Jena, Institute of Applied Physics, Albert Einstein-Str. 15, 07745 Jena, Germany}

\affil[*]{These authors contributed equally to this work.;jonas.junker@uni-jena.de}
\affil[$\dagger$]{These authors contributed equally to this work.;dolee@dtu.dk}
\affil[$\ddagger$]{jsne@fysik.dtu.dk}

\begin{abstract}
Efficient and reproducible squeezed-light sources are essential for quantum information processing and precision metrology. Compact linear standing-wave optical parametric oscillators (OPOs) are attractive because they combine low optical loss, low pump-power requirements, and large longitudinal mode spacing. In doubly resonant cavities, however, the nonlinear interaction is not determined solely by bulk phase matching: forward- and backward-generated fields recombine coherently, making the effective gain sensitive to crystal-edge termination, wavelength-dependent coating phases, and the cavity resonance condition. Here, we show that these microscopic phase contributions can produce large threshold variations between nominally similar OPOs. We combine double-pass second-harmonic generation with OPO threshold measurements to extract the relevant crystal-cavity phases and analyse three linear OPO systems. The observed devices exhibit threshold variations of up to nearly six-fold, traced to the phase-dependent nonlinear-gain envelope at accessible doubly resonant operating points. Our results establish a phase-aware framework for compact linear OPOs and provide design guidelines for reproducible low-threshold squeezed-light sources in scalable photonic quantum systems.
\end{abstract}

\setboolean{displaycopyright}{false} 

\begin{document}

\maketitle

\section{Introduction}

Squeezed states of light are a central non-classical resource in quantum optics, characterized by reduced quantum fluctuations in one field quadrature below the vacuum level~\cite{Walls1989}. This property enables measurements beyond the shot noise limit and has led to major advances in precision sensing, most prominently in interferometric gravitational-wave detection~\cite{PhysRevX.13.041021,SCHNABEL20171}. Beyond metrology, squeezed light is a key resource for continuous-variable quantum information processing, including entanglement generation, quantum teleportation, measurement-based quantum computation, and Gaussian boson sampling~\cite{RevModPhys.77.513,PhysRevLett.112.120504,PhysRevLett.119.170501}. It also underpins measurement-based protocols for generating non-Gaussian resource states, such as Gottesman--Kitaev--Preskill states~\cite{PhysRevA.64.012310,Takase2023GKP,Larsen2025IntegratedGKP}. These applications require squeezed-light sources that are efficient, stable, reproducible, and scalable.

\Acp{OPO} based on cavity-enhanced spontaneous parametric down-conversion constitutes one of the most widely used platforms for generating squeezed states of light~\cite{Wu1986SqueezedStates,SCHNABEL20171, Arnbak:19}. In these systems, a $\chi^{(2)}$ nonlinear interaction inside an optical resonator provides efficient parametric amplification, while the cavity defines well-controlled spatial and spectral modes. This combination enables high-purity squeezed states with large squeezing factors and has made \acp{OPO} a standard tool in quantum optics experiments. Recent developments have increasingly favored compact cavity implementations that enhance nonlinear gain while reducing optical loss, system size, and pump-power requirements~\cite{Arnbak:19}. Integrated waveguide sources provide an important route towards scalable squeezing~\cite{Ha2026Waveguide12dB}, but bulk \acp{OPO} remain widely used because of their low loss, excellent mode quality, experimental flexibility, and well-established theoretical and experimental understanding.

Among bulk implementations, linear standing-wave \acp{OPO}, including monolithic and hemilithic geometries, are particularly attractive. Compared with longer travelling-wave or bow-tie cavities, they can offer lower optical loss, compact form factors, reduced threshold powers, and larger longitudinal mode spacings~\cite{vahlbruch2016Detection15DB,Larsen2025IntegratedGKP}. The latter is especially useful in photon-counting-based and non-Gaussian experiments, where well-separated cavity resonances simplify spectral filtering and mode selection~\cite{NeergaardNielsen2006OddPhoton,liu2025robust,yu2026extensible}. These advantages make compact linear \acp{OPO} a promising architecture for scalable squeezed-light and photonic quantum-information experiments.

Despite these advantages, linear standing-wave \acp{OPO} exhibit an intrinsic phase sensitivity that cannot be captured by simple single-pass descriptions of the nonlinear interaction. Because the nonlinear medium is traversed in both propagation directions, the forward- and backward-generated fields recombine coherently inside the cavity~\cite{Ahlrichs2019,poveda-hospital2025ParametricProcessesNonlinear}. The effective nonlinear gain therefore depends not only on the bulk quasi-phase-matching condition, but also on the relative phases set by the termination of the poling pattern at the crystal facets, wavelength-dependent coating phases, Gouy phase shifts, and the cavity resonance condition~\cite{Paschotta1994NonlinearModeCoupling,Lastzka2007,Takeno2010,yonezawa2010GenerationSqueezedLight,Schonbeck2018,hagemann202410dBSqueezeLaser}. In practical periodically poled crystals, small uncertainties in the final domain length or boundary position can therefore translate into substantial variations in the available nonlinear gain~\cite{poveda-hospital2025ParametricProcessesNonlinear}. These phase contributions can reshape the temperature-dependent nonlinear-gain envelope and reduce the effective nonlinear efficiency.

This phase sensitivity imposes an additional practical constraint on linear \acp{OPO} operated under double resonance. In this case, the operating temperature must satisfy the simultaneous resonance condition for the fundamental and second-harmonic fields, restricting operation to a discrete set of accessible doubly resonant points. Since the microscopic phase contributions modify both the nonlinear-gain envelope and the double-resonance condition, the accessible doubly resonant points generally do not coincide with the envelope maximum. The lowest-threshold operating point can therefore differ substantially from the nominal bulk phase-matching temperature. As a result, nominally similar linear \acp{OPO} may exhibit different temperature dependencies and substantially different oscillation thresholds. A predictive description of practical linear \acp{OPO} must therefore connect the phase-sensitive nonlinear interaction, the discrete double-resonance condition, and the measurable oscillation threshold.

In this work, we investigate phase-sensitive nonlinear interactions in compact linear standing-wave \acp{OPO} based on periodically poled crystals. We develop a physically transparent model that connects single-pass, double-pass, and cavity-enhanced configurations, revealing how crystal-edge and coating phases reshape both the nonlinear-gain envelope and the accessible doubly resonant operating points. Experimentally, we combine double-pass second-harmonic generation, which isolates the phase contribution associated with the highly reflecting crystal facet, with OPO threshold measurements to reconstruct the full crystal-cavity phase response. Applying this approach to three nominally similar OPO systems, we show that microscopic phase variations can account for large threshold differences between devices. These results establish a phase-aware framework for understanding, diagnosing, and designing reproducible low-threshold squeezed-light sources for multi-source photonic quantum experiments.


\section{Theory}
\subsection{Single-pass nonlinear interaction}
\label{subsec:single_pass}

\begin{figure}[h]
    \centering
    \includegraphics[width=0.9\linewidth]{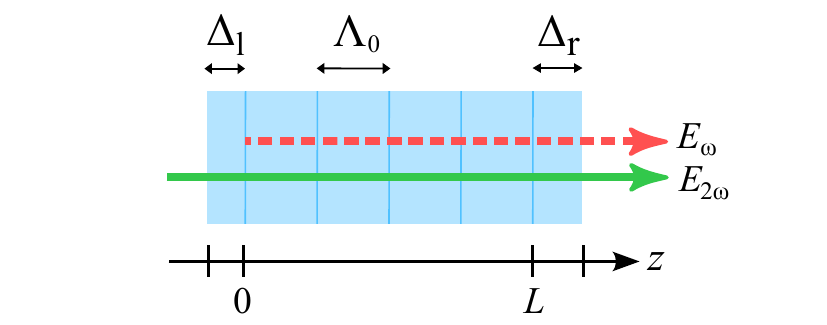}
    \caption{Single-pass nonlinear interaction inside a periodically poled crystal with two participating electric fields $E_\omega$ and $E_{2\omega}$. The poling period is given by $\Lambda_0$. In general, the crystal facets can have arbitrary termination lengths $\Delta_{\mathrm{l}}$ and $\Delta_{\mathrm{r}}$.}
    \label{fig:1a}
\end{figure}

We consider a type-0, frequency-degenerate process in a periodically poled crystal of length $L$. As illustrated in~\cref{fig:1a}, a fundamental field $E_\omega$ and a second-harmonic field $E_{2\omega}$ propagate along the crystal $z$-direction. The corresponding vacuum wavelengths are $\lambda_\omega$ and $\lambda_{2\omega}$, with angular frequencies $\omega = 2\pi c/\lambda_\omega$ and $2\omega = 2\pi c/\lambda_{2\omega}$, respectively. The crystal is periodically poled with period $\Lambda$ and may exhibit non-ideal poling termination at the input and output facets, represented by termination lengths $\Delta_{\mathrm{l}}$ and $\Delta_{\mathrm{r}}$. The electric field at frequency $\Omega_j \in \{\omega,2\omega\}$ is written in terms of a slowly varying envelope $A_j(z)$ as
\begin{equation}
    E_j(z,t)
    =
    \frac{1}{2}
    \left[
        A_j(z)e^{i(k_j z-\Omega_j t)}
        +
        A_j^*(z)e^{-i(k_j z-\Omega_j t)}
    \right].
\end{equation}
The corresponding wavevector is
\begin{equation}
    k_j(T)
    =
    \frac{n_z(\lambda_j,T)\,\Omega_j}{c},
    \qquad j \in \{\omega,2\omega\},
\end{equation}
where $n_z(\lambda_j,T)$ is the refractive index along the crystal $z$-axis at wavelength $\lambda_j$ and temperature $T$.

The refractive index is temperature dependent due to the thermo-optic effect, while thermal expansion of the crystal lattice modifies the quasi-phase-matching condition. For bulk potassium titanyl phosphate (KTiOPO$_4$, KTP), the refractive index along the crystal $z$-axis at \SI{25}{\celsius} is described by the Sellmeier equation~\cite{fradkin1999TunableMidinfraredSource}

\begin{equation}
    n_z^2(\lambda,\SI{25}{\celsius})
    =
    A
    + \frac{B}{1 - C/\lambda^{2}}
    + \frac{D}{1 - E/\lambda^{2}}
    - F\lambda^{2}.
\end{equation}

The temperature-dependent change in the refractive index is modeled by a polynomial expansion around \SI{25}{\celsius}~\cite{emanueli2003TemperaturedependentDispersionEquations},

\begin{eqnarray}
    \Delta n_z(\lambda,T)
    &=&
    (T-25)\left(
        a_{10}
        + \frac{a_{11}}{\lambda}
        + \frac{a_{12}}{\lambda^{2}}
        + \frac{a_{13}}{\lambda^{3}}
    \right)
    \nonumber \\
    &&\quad
    + (T-25)^2\left(
        a_{20}
        + \frac{a_{21}}{\lambda}
        + \frac{a_{22}}{\lambda^{2}}
        + \frac{a_{23}}{\lambda^{3}}
    \right),
\end{eqnarray}
where $T$ is expressed in degrees Celsius and the coefficients $a_{mn}$ are experimentally determined material parameters for KTP. The temperature-dependent refractive index is then given by
\begin{equation}
    n_z(\lambda,T)
    =
    n_z(\lambda,\SI{25}{\celsius})
    +
    \Delta n_z(\lambda,T).
\end{equation}
Efficient nonlinear interaction among collinear fields is achieved by quasi-phase matching, which compensates the intrinsic phase mismatch between the interacting waves. For second-harmonic generation, the bulk phase mismatch between the fundamental and second-harmonic fields is
\begin{equation}
    \Delta k(T) = k_{2\omega}(T) - 2k_{\omega}(T).
\end{equation}
In a periodically poled crystal, this mismatch is compensated by the reciprocal lattice vector of the poling structure. The residual quasi-phase mismatch is therefore given by~\cite{boyd2008NonlinearOptics}
\begin{equation}
    \Delta k_Q(T) = \Delta k(T) - \frac{2\pi}{\Lambda(T)}.
\end{equation}
Quasi-phase matching is achieved when $\Delta k_Q(T)=0$.

The poling period is weakly temperature dependent due to thermal expansion of the crystal lattice along the propagation direction. To first order, we write
\begin{equation}
    \Lambda(T)
    =
    \Lambda_0
    \left[
        1 + \alpha_z (T-T_0)
    \right],
\end{equation}
where $\Lambda_0$ is the poling period specified at the reference temperature $T_0$, and $\alpha_z$ is the linear thermal expansion coefficient along the crystal $z$-axis~\cite{smith2016thermal}. Temperature tuning of the quasi-phase-matching condition therefore arises from both the thermo-optic dependence of the refractive indices and the thermal expansion of the poling period.

For first-order quasi-phase matching, the first Fourier component of the periodically poled nonlinear coefficient gives
\begin{equation}
    d_{\mathrm{eff}} = \frac{2}{\pi} d_{33}.
\end{equation}
Here, $d_{33}$ denotes the nonlinear tensor element of KTP corresponding to the type-0 interaction. In the following, we retain only this first-order quasi-phase-matched contribution.

In second-harmonic generation, the field at frequency $2\omega$ is driven by the nonlinear polarization generated through the $\chi^{(2)}$ interaction~\cite{boyd2008NonlinearOptics}. For a type-0 process in a periodically poled crystal, this polarization can be written as
\begin{equation}
    P^{(2)}(z;2\omega)
    =
    2\varepsilon_0\, d_{\mathrm{eff}}(z)\,
    A_\omega^2(z)\, e^{i\Delta k z},
    \label{eq:NLpol}
\end{equation}
where $d_{\mathrm{eff}}(z)$ includes the spatial modulation of the nonlinear coefficient due to periodic poling. Under the slowly varying envelope approximation, Maxwell's equations yield
\begin{equation}
    \frac{dA_{2\omega}}{dz}
    =
    \kappa\, d_{\mathrm{eff}}(z)\,
    A_\omega^2\, e^{i\Delta k z},
    \label{eq:SVEA}
\end{equation}
where $\kappa$ contains the frequency, refractive-index, and normalization factors. Keeping only the first-order quasi-phase-matched Fourier component of $d_{\mathrm{eff}}(z)$ gives the effective coefficient $d_{\mathrm{eff}}$ and replaces the bulk mismatch $\Delta k$ by the residual mismatch $\Delta k_\mathrm{Q}$. For simplicity, the following treatment assumes plane-wave propagation. In practical focused-beam systems, spatial focusing modifies the exact phase-matching response according to the Boyd–Kleinman theory~\cite{Boyd1968FocusedGaussian}, but does not qualitatively change the phase-sensitive interference effects discussed here. Assuming an undepleted fundamental field, so that $A_\omega$ remains constant, the generated second-harmonic field at the output of a periodically poled crystal of length $L$ with ideal edges $(\Delta_{\mathrm{l},\mathrm{r}}=0)$ is obtained by integrating~\eqref{eq:SVEA}:
\begin{eqnarray}\nonumber
    A_{2\omega}
    &=& A_0
    \int_0^L e^{i\Delta k_\mathrm{Q} z}\,dz \\\nonumber
    &=& A_0 \frac{1}{i\Delta k_\mathrm{Q}}
    \left(e^{iL\Delta k_\mathrm{Q}}-1\right)
    \nonumber \\
    &=& A_0
    L\,e^{iL\Delta k_\mathrm{Q}/2}\,
    \mathrm{sinc}\!\left(\frac{L\Delta k_\mathrm{Q}}{2}\right).
    \label{eq:SH_field_single_pass}
\end{eqnarray}
Here, $A_0$ includes the constant factors, including $d_{\mathrm{eff}}$, normalization constants, and the undepleted fundamental amplitude. The phase factor $e^{iL\Delta k_\mathrm{Q}/2}$ is a global phase that depends on the choice of coordinate origin, while the sinc term describes the usual temperature-dependent phase-matching envelope. Here, we use $\Delta k_\mathrm{Q}$ for the quasi-phase-matched nonlinear interaction in the poled region, while $\Delta k$ is retained for the propagation phase accumulated in the residual edge regions. Perfectly constructive interference between successive domains occurs when $\Delta k_\mathrm{Q}=0$. We note that the poled interaction length is given by $L = 405\Lambda \approx \SI{10.0}{\milli\meter}$ for the type-0 interaction considered here, corresponding to a poling period of $\Lambda = \SI{24.7}{\micro\meter}$ specified for phase matching near \SI{35}{\celsius}.

\eqref{eq:SH_field_single_pass} remains a good approximation for crystals with finite edge regions $(\Delta_{\mathrm{l},\mathrm{r}}\neq 0)$, provided that $\Delta_{\mathrm{l},\mathrm{r}}\lesssim\Lambda$ and $\Delta_{\mathrm{l},\mathrm{r}}\ll L$. In single-pass propagation, the bulk contribution then dominates the generated field, so the edge regions have only a negligible effect on the output amplitude. As shown below, however, the phases associated with these edge regions become important when forward- and backward-generated fields interfere in double-pass and cavity-enhanced configurations.


\subsection{Double-pass nonlinear interaction}
\label{subsec:double_pass}
\begin{figure}[h]
    \centering
    \includegraphics[width=0.9\linewidth]{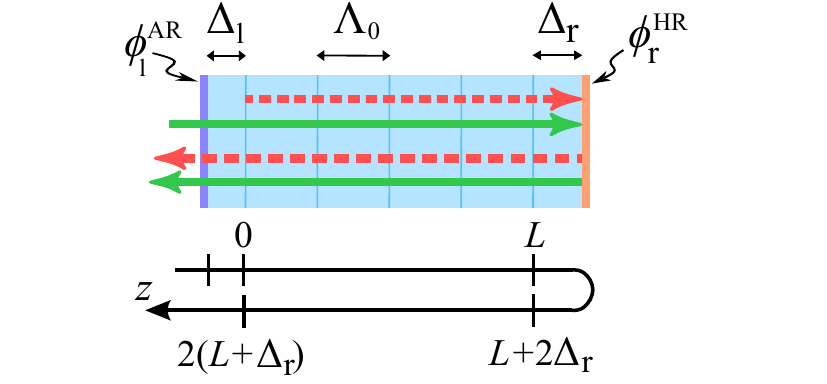}
    \caption{Double-pass nonlinear interaction in a crystal with anti-reflective (AR) and high-reflective (HR) surfaces at which both fields are back-reflected.} 
    \label{fig:1b}
\end{figure}
A straightforward way to enhance the nonlinear conversion efficiency is to effectively increase the interaction length of the nonlinear medium. One practical implementation is a double-pass configuration of the nonlinear crystal, as illustrated in~\cref{fig:1b}. In this scheme, the crystal is typically coated with an \ac{AR} coating on one surface and an \ac{HR} coating on the opposite surface, for both participating wavelengths, such that the fundamental field traverses the crystal twice. For \ac{SHG}, the generated second-harmonic field in a double-pass configuration corresponds to the coherent sum of two single-pass contributions arising from forward and backward propagation through the crystal,
\begin{equation}
    A^{\mathrm{SHG}}_{2\omega}
    \propto
    \int_{0}^{L} e^{i\Delta k_\mathrm{Q} z}\,dz
    +
    e^{i \Phi_{\mathrm{R}}}
    \int_{L+2\Delta_{\mathrm{r}}}^{2(L+\Delta_{\mathrm{r}})}
    e^{i\Delta k_\mathrm{Q} z}\,dz.
    \label{eq:SH_field_double_pass}
\end{equation}

\cref{eq:SH_field_double_pass} provides an accurate and physically transparent description of the double-pass \ac{SHG} process. In this model, contributions from fields generated at the crystal edges are neglected, as their effect on the total second-harmonic field is small. Thus, we can define the phase $\Phi_{\mathrm{R}}$ denotes the effective relative phase of the backward-generated second-harmonic contribution with respect to the forward-generated contribution. It includes the relative reflection phase at the high-reflectivity (HR) coated crystal surface, $\phi^{\text{HR}}_{\mathrm{r}}$, defined with respect to the corresponding nonlinear driving field, as well as the residual propagation phase associated with the right crystal edge, $\psi_{\mathrm{r}} = 2\Delta_{\mathrm{r}}\Delta k$. With this convention, the total phase is given by $\Phi_{\mathrm{R}} = \phi^{\text{HR}}_{\mathrm{r}} + \psi_{\mathrm{r}}$.

\begin{figure}[h]
    \centering
    \includegraphics[width=0.9\linewidth]{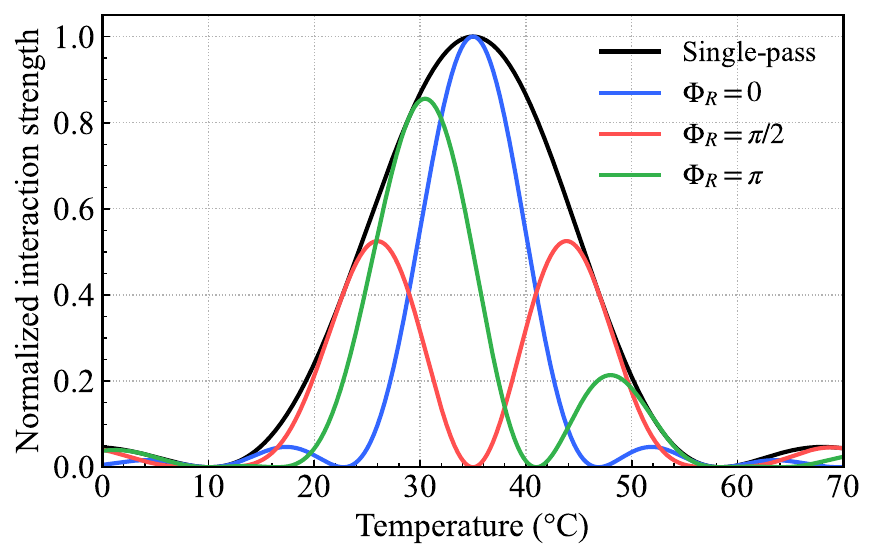}
    \caption{Comparison of the nonlinear interaction strength for the single-pass and double-pass case with different phases $\Phi_\mathrm{R}$. Plotting parameters:  $\lambda_{\omega}=\SI{1550}{\nano \metre}$, $\lambda_{2\omega}=\SI{775}{\nano\metre}$, $\Lambda_0=\SI{24.7}{\micro \metre}$, $L=405 \times \Lambda_0\approx \SI{1}{\centi \metre}$.} 
    \label{fig:SPDBcomparison}
\end{figure}

It is insightful to compare single-pass and double-pass nonlinear interactions. \cref{fig:SPDBcomparison} shows the temperature-dependent second-harmonic power $P^\text{SHG}_{2\omega} \propto \abs{A^\text{SHG}_{2\omega}}^2$ generated by pumping the crystal at the fundamental frequency. All traces are normalised to the highest possible single-pass and double-pass power, respectively. The black trace corresponds to single-pass propagation and exhibits the characteristic sinc-like phase-matching response. The blue trace shows the double-pass configuration for a relative phase $\Phi_\text{R}=0$, which is equivalent to single-pass conversion in a crystal of length $2L$. For $\Phi_\text{R}\neq0$, as illustrated by the red and green traces, the height and positions of maxima and minima shift with temperature. In particular, for $\Phi_\text{R}=\pi$, the second-harmonic power vanishes at the nominal phase-matching temperature $T_0$ due to destructive interference between the forward- and backward-generated contributions. This interference-controlled regime is unique to the double-pass and does not appear in single-pass operation. 

\subsection{Nonlinear interaction in a linear cavity}
\label{subsec:cavity}
\begin{figure}[h]
    \centering
    \includegraphics[width=0.9\linewidth]{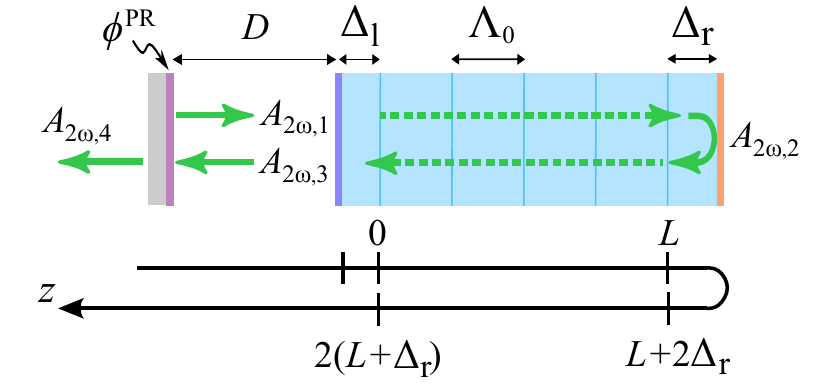}
    \caption{Fundamental and second-harmonic fields in a linear cavity. The circulating cavity field contains forward- and backward-propagating components whose relative phases depend on the facet termination lengths $\Delta_{\mathrm{l}}$ and $\Delta_{\mathrm{r}}$.}
    \label{fig:1c}
\end{figure}

To further enhance the interaction strength, the nonlinear crystal is often placed inside an optical cavity. While traveling-wave cavities can be employed \cite{junker2022FrequencyDependentSqueezingDetuned}, we focus here on standing-wave cavities, which are widely used in quantum optical experiments \cite{vahlbruch2016Detection15DB}. In this configuration, the cavity is typically formed by a high-reflectivity coating on one face of the crystal and a partially reflective mirror separated by a distance $D$ from the opposite face, as shown in~\cref{fig:1c}. This mirror serves as the input-output coupler of the cavity. In monolithic or hemilithic implementations, the input-output coupler may also be realized directly on the crystal surface, without qualitatively changing the following analysis. Its partial reflective side is characterized by amplitude reflectivity $\rho$ and transmissivity $\tau$. The anti-reflective side is assumed to have a transmission of 1. We consider a doubly resonant linear cavity, in which both the fundamental and second-harmonic fields satisfy a cavity resonance condition. Efficient nonlinear interaction in such a system requires simultaneous fulfillment of the nonlinear phase-matching condition and resonance of both wavelengths. 

The single-pass phase-matching condition can be extended to a cavity configuration by introducing effective intracavity field relations. We define the intracavity second-harmonic fields $A_{2\omega,1}$, $A_{2\omega,2}$, and $A_{2\omega,3}$, together with the output field $A_{2\omega,4}$. These fields are connected through the cavity boundary conditions, together with the second-harmonic fields generated during the forward and backward passes through the poled region. The phases of the generated fields are defined relative to the corresponding fundamental fields. Following the effective phase convention introduced in the double-pass SHG model, the residual propagation and coating reflection phases are incorporated into the effective relative phases $\Phi_{\mathrm{R}}$ and $\Phi_{\mathrm{L}}$, which describe the total phase accumulated by the generated second-harmonic fields with respect to the corresponding fundamental fields during one cavity round-trip. For the left side of the cavity in Fig~\ref{fig:1c}, the effective phase is given by $\Phi_{\mathrm{L}}=\phi^{\mathrm{PR}}_{\mathrm{l}}+2\phi^{\mathrm{AR}}_{\mathrm{l}}+\psi_{\mathrm{l}}$, where $\phi^{\mathrm{PR}}_{\mathrm{l}}$ denotes the reflection phase of the PR-coated input coupler, $\phi^{\mathrm{AR}}_{\mathrm{l}}$ the transmission phase of the AR-coated crystal facet, and $\psi_{\mathrm{l}}$ the residual propagation phase associated with the left crystal edge. The cavity field relations can therefore be written as
\begin{align}
    A_{2\omega,1} &= \rho\, A_{2\omega,3}, \\
    A_{2\omega,2} &= e^{-i k_{2\omega} D}
    \left(
    A_{2\omega,1}
    + A_{2\omega}^{\mathrm{r}}
    e^{-i\Phi_{\mathrm{R}}}
    \right), \\
    A_{2\omega,3} &= e^{-i k_{2\omega} D}
    \left(
    A_{2\omega,2}
    + A_{2\omega}^{\mathrm{l}}
    e^{-i\Phi_{\mathrm{L}}}
    \right), \\
    A_{2\omega,4} &= i\tau\, A_{2\omega,3}.
\end{align}

Here, $A_{2\omega}^{\mathrm{l}}$ and $A_{2\omega}^{\mathrm{r}}$ denote the second-harmonic fields generated during the forward and backward passes through the poled region, respectively, with
\begin{align}
    A_{2\omega}^{\mathrm{r}}
    &\propto
    \int_{0}^{L}
    e^{i\Delta k_\mathrm{Q} z}\,dz, \\
    A_{2\omega}^{\mathrm{l}}
    &\propto
    \int^{2(L+\Delta_{\mathrm{r}})}_{L+2\Delta_{\mathrm{r}}}
    e^{i\Delta k_\mathrm{Q} z}\,dz.
\end{align}
Solving this system yields an expression for the output field that depends on the cavity round-trip phase, the mirror reflectivity, and the effective relative phases accumulated by the generated fields:
\begin{equation}
    A_{2\omega,4}
    =
    \frac{
    i\,\tau
    \left(
        A_{2\omega}^{\mathrm{l}}
        e^{i\left(k_{2\omega}D+\Phi_{\mathrm{L}}\right)}
        +
        A_{2\omega}^{\mathrm{r}}
        e^{i\Phi_{\mathrm{R}}}
    \right)
    }{
    e^{2ik_{2\omega}D}-\rho
    },
    \label{eq:field_cavity}
\end{equation}
To isolate the envelope of the cavity-enhanced phase-matching response from the cavity double-resonance condition, we choose the global cavity phase as the reference phase, corresponding to a resonant second-harmonic cavity field. The remaining phase-sensitive behavior is therefore governed by the effective relative phases $\Phi_{\mathrm{L}}$ and $\Phi_{\mathrm{R}}$, yielding
\begin{equation}
    A_{2\omega}^{\mathrm{env}}(T,\Phi_{\mathrm{L}},\Phi_{\mathrm{R}})
    =
    A_{2\omega,4}
    =
    \frac{
    i\,\tau
    \left(
        A_{2\omega}^{\mathrm{l}}
        e^{i\Phi_{\mathrm{L}}}
        +
        A_{2\omega}^{\mathrm{r}}
        e^{i\Phi_{\mathrm{R}}}
    \right)
    }{
    1-\rho
    },
    \label{eq:envelope_cavity}
\end{equation}

For the threshold fitting in \cref{subsec:linearOPOexperiment}, we use the corresponding normalized nonlinear interaction strength, defined as the squared modulus of the envelope field relative to its optimum value,
\begin{equation}
    \eta_{\mathrm{nl}}(T,\Phi_{\mathrm{L}},\Phi_{\mathrm{R}})
    =
    \frac{
    \left|
    A_{2\omega}^{\mathrm{env}}
    (T,\Phi_{\mathrm{L}},\Phi_{\mathrm{R}})
    \right|^2
    }{
    \left|
    A_{2\omega}^{\mathrm{env}}
    \right|_{\mathrm{opt}}^2
    },
    \label{eq:eta_nl}
\end{equation}

Next, we determine the doubly resonant operating points for the fundamental and second-harmonic fields inside the cavity. The geometrical cavity round-trip length is given by $L_{\mathrm{cav}} = 2(D + \Delta_{\mathrm{l}} + L + \Delta_{\mathrm{r}})$, which includes the propagation through the crystal, the edge regions, and the free-space section. The accumulated relative phase per round trip between the second-harmonic field and twice the fundamental field inside the crystal is
\begin{eqnarray}\nonumber
    \Phi_{\mathrm{rt}}(T)
    &=&
    2L\!\left[
        k_{2\omega}(T)-2k_{\omega}(T)
    \right]
    +
    \Phi_{\mathrm{L}}
    +
    \Phi_{\mathrm{R}}\\
    &+&
    2D\!\left[
        k_{2\omega,0}-2k_{\omega,0}
    \right],
    \label{eq:double_resonance}
\end{eqnarray}
where $k_{j,0}=\Omega_j/c$ denotes the free-space wavevector. For exact frequency doubling in free space, $k_{2\omega,0}=2k_{\omega,0}$, such that the free-space contribution vanishes. Doubly resonant operation occurs when the relative round-trip phase satisfies $\Phi_{\mathrm{rt}} \bmod 2\pi = 0$.
\begin{figure*}[t]
  \centering
  \includegraphics[width=1.0\linewidth]{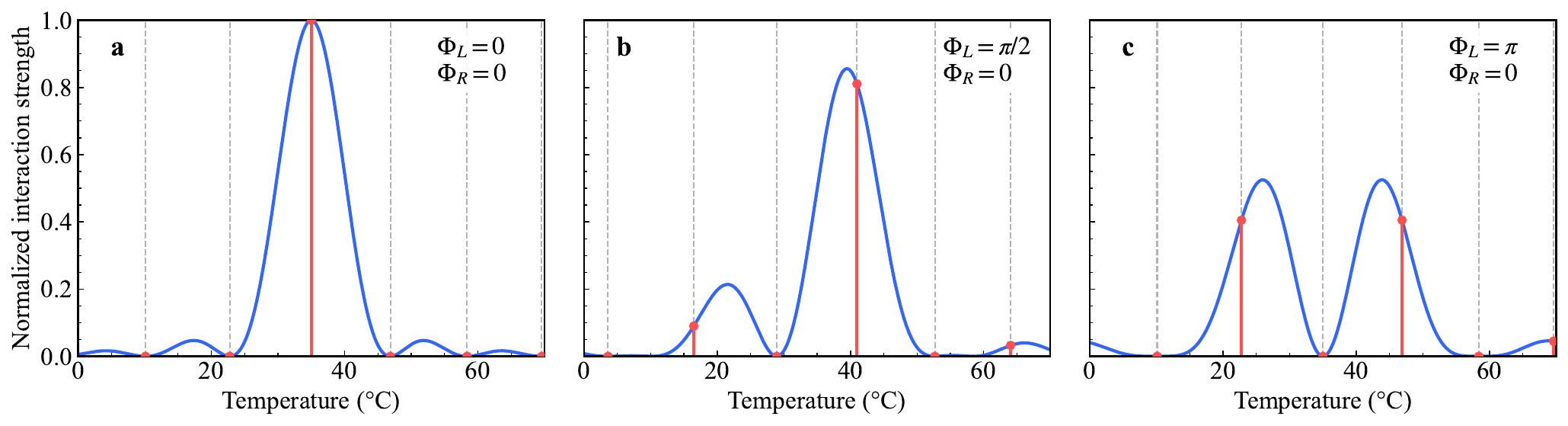}
  \caption{Phase-sensitive nonlinear interaction in a doubly resonant linear cavity for representative phase combinations $\Phi_L$ and $\Phi_R$. The blue curves show the cavity-enhanced nonlinear interaction envelope, while the gray dashed lines indicate the discrete doubly resonant operating points allowed by the cavity resonance condition. Depending on the phase contributions from the crystal-edge regions, the accessible operating points can strongly suppress the achievable nonlinear gain.}
  \label{fig:cavity_comparison}
\end{figure*}

To illustrate the intracavity phase-sensitive dynamics, \cref{fig:cavity_comparison} shows the normalized nonlinear interaction strength for representative phase combinations $\Phi_{\mathrm{L}}$ and $\Phi_{\mathrm{R}}$. The blue curves represent the cavity-enhanced nonlinear-interaction envelope, evaluated with the second-harmonic cavity response on resonance, before imposing the simultaneous double-resonance condition. In practice, however, simultaneous resonance of the fundamental and second-harmonic fields occurs only at discrete temperatures, indicated by the gray dashed lines. The corresponding accessible interaction strengths are marked in red.

For the ideal case $\Phi_\mathrm{L}=\Phi_\mathrm{R}=0$, the cavity envelope coincides with the double-pass response discussed in \cref{subsec:double_pass}. In this situation, one cavity round trip corresponds to the same effective interaction length as in the double-pass configuration, while the forward- and backward-generated fields acquire no additional relative phase. Consequently, the nonlinear interaction remains fully constructive near the phase-matching temperature.

When the phase contributions from the two crystal-edge regions differ, both the envelope response and the locations of the doubly resonant operating points are modified. As a result, the experimentally accessible operating temperatures may no longer coincide with the maxima of the nonlinear interaction envelope. For example, for $\Phi_\mathrm{L}=\pi$ or $\Phi_\mathrm{L}=\pi/2$ with $\Phi_\mathrm{R}=0$, the doubly resonant operating points shift close to minima of the nonlinear interaction envelope, such that the cavity-enhanced nonlinear interaction becomes strongly suppressed over a broad temperature interval between roughly \SI{25}{\celsius} and \SI{45}{\celsius}. This demonstrates that, in practical linear OPO systems, the achievable nonlinear gain is jointly determined by the cavity resonance condition and the phase-sensitive interference between forward- and backward-generated fields.

\begin{figure}[h]
    \centering  
    \includegraphics[width=1.0\linewidth]{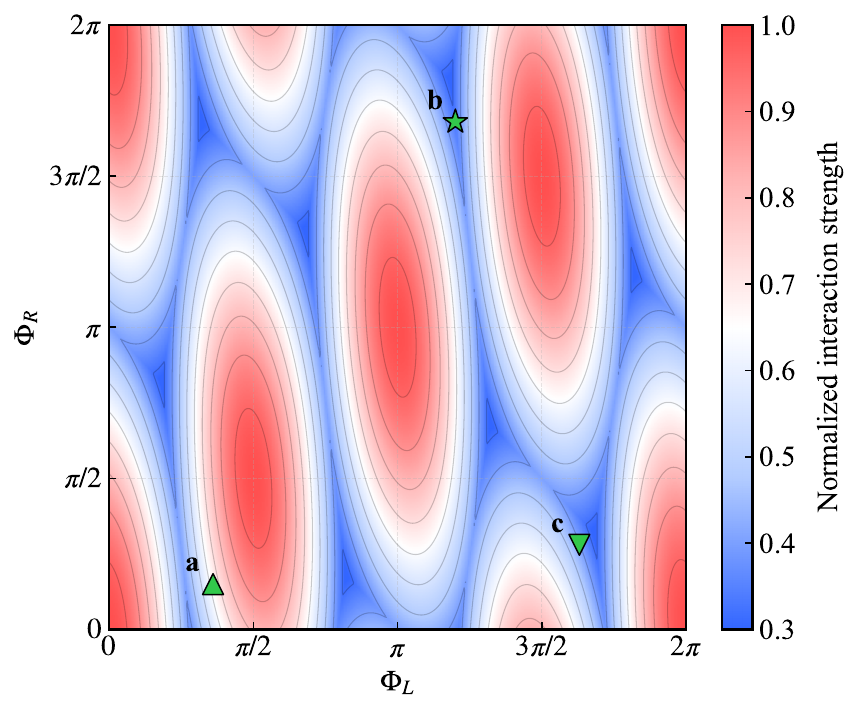}
    \caption{Maximum reachable normalized nonlinear interaction strength as a function of the phase contributions $\Phi_{\mathrm{L}}$ and $\Phi_{\mathrm{R}}$, optimized over the accessible doubly resonant operating temperatures. The interaction strength is bounded from below by approximately $0.29$, corresponding to a maximum threshold penalty of $1/0.29 \simeq 3.44$ under temperature-optimized operation. The green markers indicate the experimentally investigated OPO systems: OPO$_{\mathrm{A}}$ (a), OPO$_{\mathrm{B}}$ (b), and OPO$_{\mathrm{C}}$ (c).}
    \label{fig:density_plot}
\end{figure}

The periodic parameter space spanned by $\Phi_{\mathrm{L}}$ and $\Phi_{\mathrm{R}}$ is shown in \cref{fig:density_plot}. For each phase combination, we determine the maximum reachable normalized nonlinear interaction strength by optimizing over the accessible doubly resonant operating temperatures. The density plot shows that, within this model, the temperature-optimized interaction strength does not fall below approximately $0.29$ of the ideal value. For the corresponding down-conversion process in an \ac{OPO}, this implies a maximum threshold penalty of $1/0.29 \approx 3.44$ relative to the ideal phase case.

\noindent

\section{Experiment}
\label{sec:experiment}
\begin{figure}[h]
    \centering
    \includegraphics[width=1.0\linewidth]{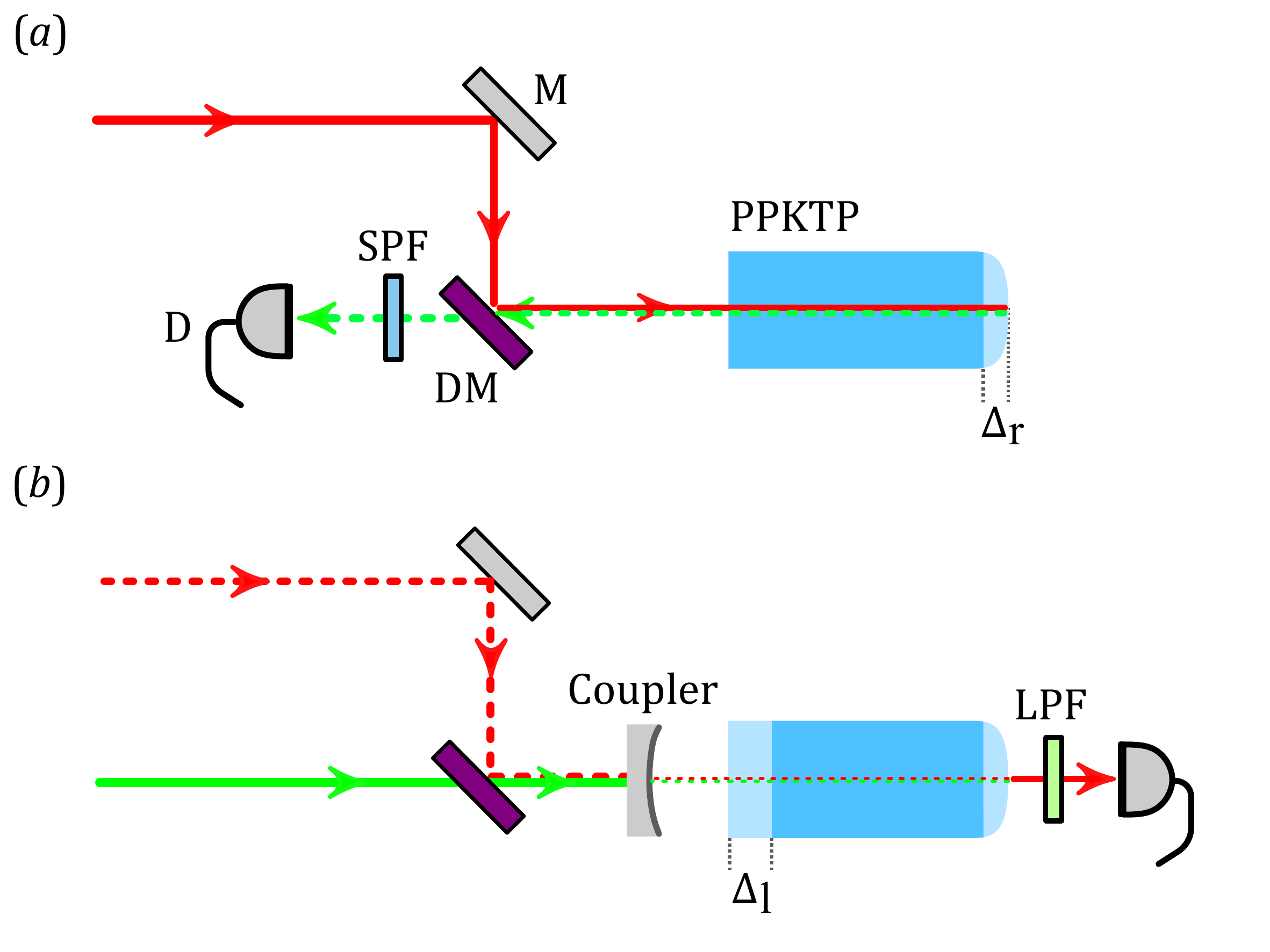}
    \caption{Experimental setups for characterizing phase contributions of the crystal. (a) Double-pass second-harmonic generation (SHG) configuration used to extract phase contributions associated with the HR-coated crystal facet. (b) Linear OPO configuration forming a standing-wave cavity with an external input coupler for threshold measurements. M: mirror, DM: dichroic mirror, SPF: short-pass filter, LPF: long-pass filter, D: detector.}
    \label{fig:setup}
\end{figure}

For a type-0 nonlinear process, we employ three periodically poled KTiOPO$_4$ (PPKTP) crystals (Raicol Inc.) with a poling period of \SI{24.7}{\micro\metre}, optimized for quasi-phase matching between \SI{1550}{\nano\metre} and \SI{775}{\nano\metre} at \SI{35}{\celsius}. To integrate the crystal into a linear OPO configuration, the crystal edges are polished asymmetrically: the left edge is plano with a flatness of $\lambda/10$, while the right edge is convex with a radius of curvature of \SI{10}{\milli\metre}. This polishing process introduces an uncertainty in the total crystal length of approximately \SI{\pm 100}{\micro\metre}. Consequently, the crystal edges do not, in general, coincide with a well-defined point of the poling pattern, so that the final domain period is effectively random. This gives rise to an effectively random phase contribution from the crystal edges. Both edges are subsequently coated with multilayer dielectric coatings (Laseroptik Inc.), providing an anti-reflection (AR) coating on the plano edge and a high-reflection (HR) coating on the convex edge for the both wavelengths. The crystal temperature is controlled using a custom-designed mount equipped with a Peltier element and a temperature sensor. A feedback loop ensures long-term temperature stability over a range from 
\SI{20}{\celsius} to \SI{60}{\celsius}. 

\subsection{Double-pass SHG experiment}
We first characterize the double-pass nonlinear response of the PPKTP crystals, by taking temperature-dependent SHG measurements. The experimental setup used for this characterisation is shown in Fig.~\ref{fig:setup}a. A continuous-wave \SI{1550}{\nano\metre} laser (NKT Photonics Inc.) is reflected by a dichroic mirror and injected into the PPKTP crystal. The frequency-doubled light generated via the double-pass SHG process exits through the plano side and is directed to a power detector placed in the transmitted output port of the dichroic mirror. The grooved structure of the mount provides a mechanical reference for normal incidence on the plano side of the crystal, which was verified by overlapping the reflected beam from the crystal surface with the incident beam.

\begin{figure*}[h]
  \centering
  \includegraphics[width=1.0\linewidth]{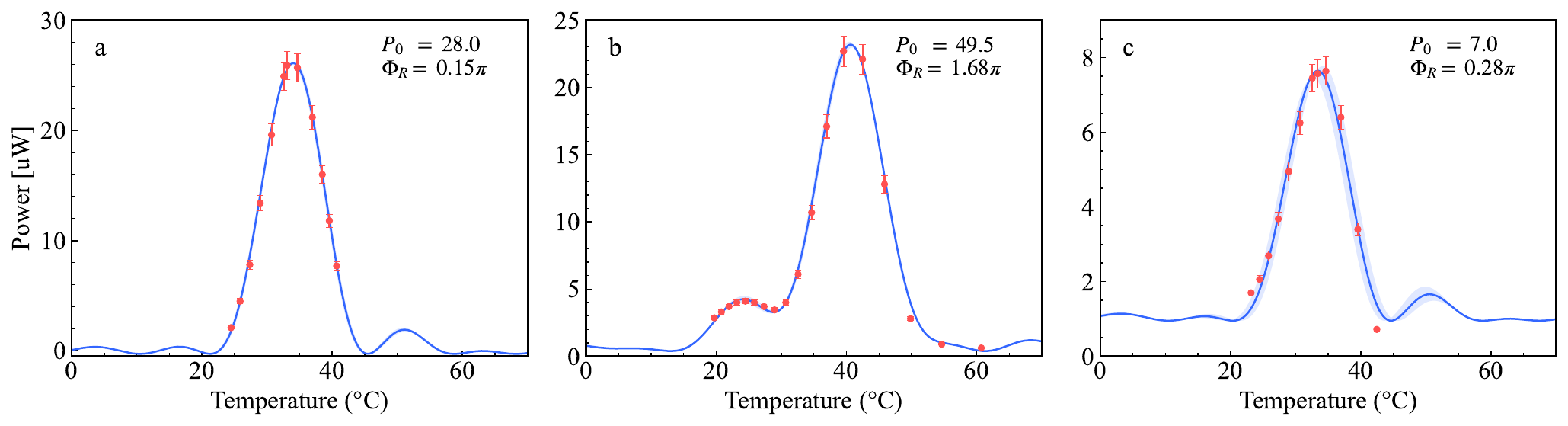}
  \caption{
  Measured double-pass SHG intensity as a function of crystal temperature for different PPKTP crystals together with fitted curves obtained from the minimization of~\eqref{eq:cost_shg}. The red points represent the measured SHG intensities with error bars corresponding to the detector uncertainty. The solid blue curves show the fitted theoretical model, while the shaded blue regions indicate the $1\sigma$ confidence interval associated with the fitted $\Phi_\mathrm{R}$ parameter. 
\textbf{a}. $\Phi_\mathrm{R} = 0.15 \pm 0.012\pi$, $\beta = 0.955$, $P_\mathrm{g} = \SI{-0.281}{\micro \watt}$. 
\textbf{b}. $\Phi_\mathrm{R} = -0.32 \pm 0.013\pi$, $\beta = 0.398$, $P_\mathrm{g} = \SI{0.393}{\micro \watt}$. 
\textbf{c}. $\Phi_\mathrm{R} = 0.28 \pm 0.043\pi$, $\beta = 0.980$, $P_\mathrm{g} = \SI{0.953}{\micro \watt}$.
}
  \label{fig:SHG_fit_comparison}
\end{figure*}

Figure~\ref{fig:SHG_fit_comparison} shows the measured SHG intensity as a function of temperature for the three different crystals. We use a model function to fit the measured data:
\begin{equation}
P^{\mathrm{SHG}}_{\mathrm{fit}}(T)
=
P_{0}
\left|
A_{2\omega}(T)
\left(
1 + \beta e^{i\Phi_R}
\right)
\right|^{2}
+
P_\mathrm{g} ,
\label{eq:fit_shg}
\end{equation}
where $A_{2\omega}(T)$ corresponds to the normalized temperature-dependent phase-matching response introduced above, with all traces normalized to the maximum achievable SHG power. The parameter $P_0$ is therefore a dimensionless scaling factor for the measured SHG intensity, while $P_\mathrm{g}$ is a constant background offset power. The parameter $\beta$ is the relative field amplitude of the backwards-generated field and accounts for imperfect spatial mode matching and unequal nonlinear conversion efficiencies between the two propagation directions. The crystal length $L$, poling period $\Lambda$, and the temperature dependence of $\Delta k_\mathrm{Q}(T)$ are fixed by the independently determined crystal parameters and material relations introduced above. The free fit parameters are therefore $\Phi_{\mathrm R}$, $\beta$, $P_0$, and $P_\mathrm{g}$. To extract the phase contribution from the HR-coated crystal facet, we minimize the weighted least-squares cost function
\begin{equation}
\mathcal{C}\left(P_0,\beta,\Phi_R,P_\mathrm{g}\right)
=
\sum_i
\left(
\frac{
P_i - P^{\mathrm{SHG}}_{\mathrm{fit}}(T_i)
}{
\sigma_{P_i}
}
\right)^2 ,
\label{eq:cost_shg}
\end{equation}
where \(P_i\) denotes the measured SHG intensity at temperature \(T_i\), and \(\sigma_{P_i}\) corresponds to the associated measurement uncertainty of $\pm5~\%$. The fitted results show that the effective relative phase accumulated at the right crystal edge determines the temperature-dependent envelope of the SHG efficiency.

The absolute SHG powers shown in Fig.~\ref{fig:SHG_fit_comparison} differ between the three measurements because the experiments were performed under slightly different operating conditions. In particular, OPO$_\text{A}$ and OPO$_\text{C}$ were measured using the same SHG configuration with different pump powers and therefore exhibit comparable fitted \(\beta\) values, whereas OPO$_\text{B}$ was measured later using an updated crystal mount configuration. Consequently, moderate variations in the fitted amplitude parameters are expected and do not affect the extraction of the phase-sensitive interference features that constitute the main focus of the present analysis. The fitted phase contributions associated with the right crystal edge are $0.15\pi$, $1.68\pi$, and $0.28\pi$. Taking into account the phase shift introduced by the HR coating at both wavelengths ($\phi_{\mathrm r}^{\mathrm{HR}} = 221.7^\circ$ (1550 nm) $- 158.9^\circ$ (775 nm) $= 0.349\pi$), obtained from the coating specifications provided by the manufacturer, these values correspond to the effective right-edge lengths of \SI{9.91}{\micro\meter}, \SI{4.07}{\micro\meter}, and \SI{11.45}{\micro\meter}, respectively.

These experimentally inferred parameters are consistent with the polished length-error range, and the model reproduces the observed temperature-dependent behaviour governed by the total phase contribution from the right crystal edge. However, the plano left edge does not contribute to the interference effects in the double-pass SHG configuration, so double-pass SHG results do not give us a full characterization of the crystal.

\begin{table}[h]
\centering
\begin{tabular}{lccc}
\hline
 & \textbf{OPO$_\text{A}$} & \textbf{OPO$_\text{B}$} & \textbf{OPO$_\text{C}$} \\
\hline

$\Phi_\mathrm{R}$ & $0.15\pi$ & $1.68\pi$ & $0.28\pi$ \\
$\Delta_{\mathrm{r}} (\SI{}{\micro\meter}$) & 9.91 & 4.07 & 11.45 \\

\hline

$\Phi_\mathrm{L}$ & $0.36\pi$ & $1.20\pi$ & $1.63\pi$ \\
$\Delta_{\mathrm{l}} (\SI{}{\micro\meter}$) & 0.20 & 10.57 & 3.53 \\

\hline

$P^\text{opt}_{\mathrm{thr}}$ (mW) & 10.8 & 23.2 & 16.1 \\

$T_\mathrm{fit}$ ($^\circ$C) & 28.8 / 40.9 & 22.8 / 35.1 & 36.1 / 48.0 \\
$P_{\mathrm{fit}}$ (mW) & 70.3 / 16.4 & 70.8 / 67.2 & 101.1 / 57.7 \\

\hline

$T_\text{res}$ ($^\circ$C) & 28.4 / 39.8 & 24.5 / 37.1 & 35.3 / 46.4 \\
$P_{\mathrm{thr}}$ (mW) & 70 / 17 & 71 / 67 & 101 / 57 \\

\hline
\end{tabular}
\caption{Summary of extracted phase contributions ($\Phi_\mathrm{L}$, $\Phi_\mathrm{R}$) and corresponding effective edge lengths ($\Delta_{\mathrm{l}}$, $\Delta_{\mathrm{r}}$). The measured OPO threshold powers ($P_{\mathrm{thr}}$) are compared with the fitted optimal values ($P^{\text{opt}}_{\mathrm{thr}}$), showing that the combined phase contributions lead to large variations in the oscillation threshold. For each crystal, the slash-separated values denote the measured oscillation thresholds at the two accessible doubly resonant temperatures. The ideal compensation lengths required to cancel the coating-induced phase shifts are $\Delta^{\mathrm{ideal}}_\mathrm{r} = \SI{4.31}{\micro \meter}$ and $\Delta^{\mathrm{ideal}}_\mathrm{l} = \SI{4.25}{\micro \meter}$.}
\label{table:summary}
\end{table}

\subsection{Linear OPO experiment}
\label{subsec:linearOPOexperiment}
For the linear OPO operation, an input coupling mirror with a radius of curvature of \SI{18}{\milli\metre} is placed in front of the crystal, as shown in Fig.~\ref{fig:setup}b. The partial-reflective (PR) coating on the crystal-facing side of the coupler, with a power reflectivity of $\mathrm{75}\,\%$ at \SI{775}{\nano\meter} and $\mathrm{94}\,\%$ at \SI{1550}{\nano\meter}, completes the standing-wave cavity. A weak \SI{1550}{\nano\metre} fundamental field is reflected on a dichroic mirror and is then sent through the AR coated plano side of this coupler. Additionally, a pump field at \SI{775}{\nano\metre} is sent through the same port into the cavity. This pump is generated by a second-harmonic generation module using the same fundamental seed laser as in the 
double-pass SHG experiment (NKT Photonics Inc.). 

\begin{figure*}[h]
  \centering
  \includegraphics[width=1.0\linewidth]{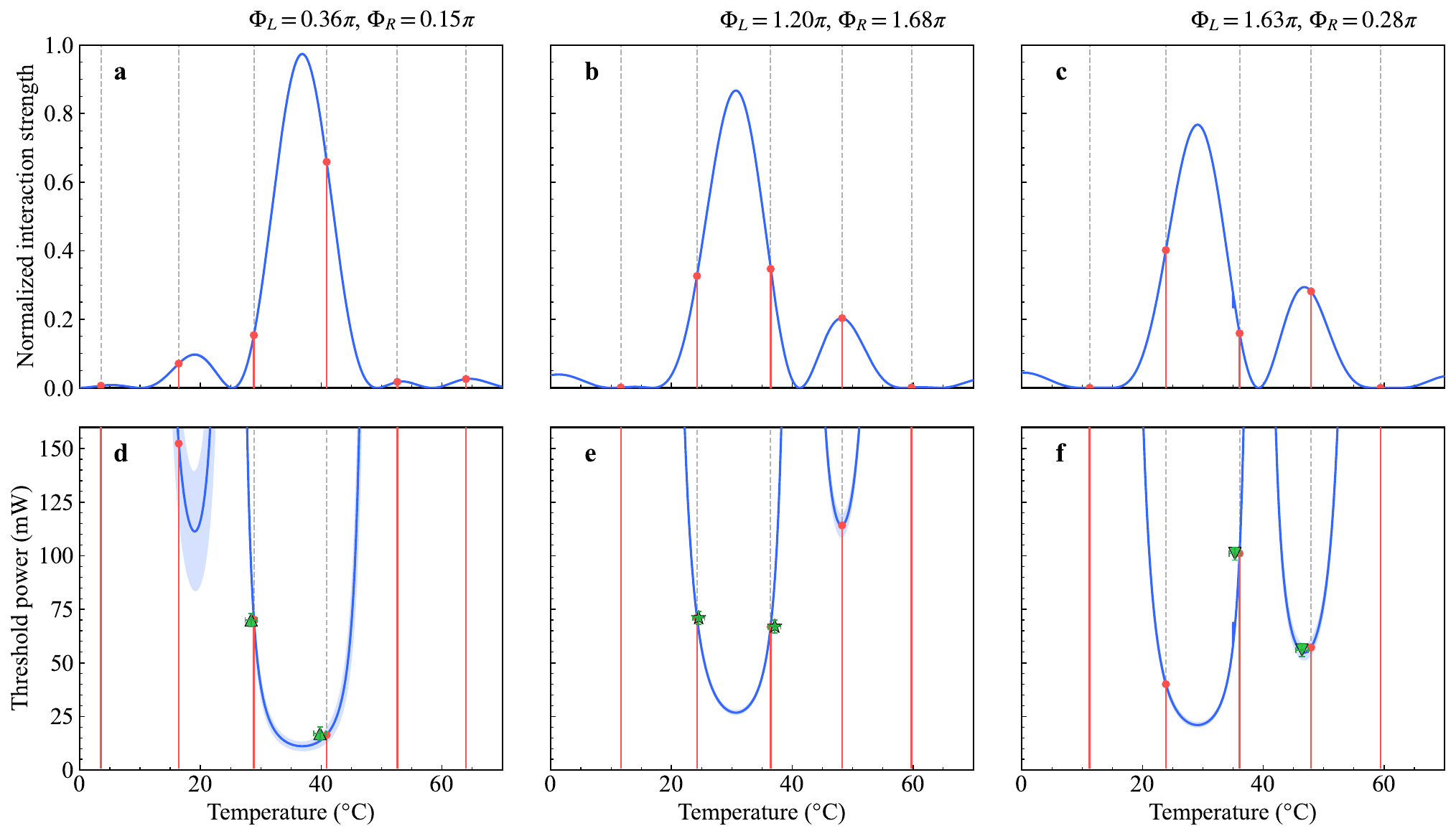}
  \caption{Normalized nonlinear interaction strength (\textbf{a--c}) and corresponding oscillation threshold power (\textbf{d--f}) for OPO$_{\mathrm{A}}$, OPO$_{\mathrm{B}}$, and OPO$_{\mathrm{C}}$, obtained from the fit based on~\eqref{eq:fit_OPO} using the experimentally extracted phase contributions summarized in \cref{table:summary}. The blue curves show the model predictions, and the red markers indicate the experimentally accessible doubly resonant operating points. The fitted optimal threshold powers are $P_{\mathrm{thr}}^{\mathrm{opt}}=\SI{10.8}{\milli\watt}$, $\SI{23.2}{\milli\watt}$, and $\SI{16.1}{\milli\watt}$ for OPO$_{\mathrm{A}}$, OPO$_{\mathrm{B}}$, and OPO$_{\mathrm{C}}$, respectively. The shaded regions indicate the accepted parameter ranges defined by $\mathcal{C}\leq\mathcal{C}_{\min}+2.30$, corresponding to $\Phi_{\mathrm{L}}/\pi=0.343$--$0.372$, $1.196$--$1.205$, and $1.625$--$1.634$, and $P_{\mathrm{thr}}^{\mathrm{opt}}=\SIrange{8.5}{13.0}{\milli\watt}$, $\SIrange{22.2}{24.3}{\milli\watt}$, and $\SIrange{14.9}{17.1}{\milli\watt}$ for OPO$_{\mathrm{A}}$, OPO$_{\mathrm{B}}$, and OPO$_{\mathrm{C}}$, respectively. The green markers indicate the experimentally investigated OPO systems: OPO$_{\mathrm{A}}$, OPO$_{\mathrm{B}}$, and OPO$_{\mathrm{C}}$ in~\cref{table:summary}.}
  \label{fig:TBD}
\end{figure*}

By scanning the cavity length, we simultaneously monitor the resonance peaks of both the fundamental \SI{1550}{\nano\metre} and second-harmonic \SI{775}{\nano\metre} fields. Due to dispersion and thermal expansion, the resonance conditions of the two wavelengths shift differently with temperature. By tuning the crystal temperature, we identify the operating points where the two resonances overlap within the accessible temperature tuning range, corresponding to doubly resonant conditions of the linear OPO. In the compact, mechanically stable OPO implementation, the crystal oven was integrated into a sealed cavity, which was not fully thermally isolated from the surrounding cavity elements, limiting the upper operating temperature to approximately \SI{50}{\celsius}. The lower operating temperature limit of approximately \SI{25}{\celsius} was chosen to avoid condensation on the crystal. Consequently, only two doubly resonant temperatures could be directly accessed experimentally. Nevertheless, comparison with the theoretical double-resonance condition~\eqref{eq:double_resonance} indicates the existence of additional resonant points outside this range, allowing the corresponding threshold behaviour to be predicted by the model.

At the doubly resonant temperatures, phase-sensitive amplification and de-amplification of the intracavity field are observed. To characterize the nonlinear response of each crystal and quantify its impact on linear OPO performance, we determined the oscillation threshold at each accessible doubly resonant temperature. The pump power was gradually increased while monitoring the \SI{1550}{\nano\metre} output from the cavity. The oscillation threshold was determined from the pump power at which the output signal started to increase sharply. 

The resulting resonance temperatures, $T_{\mathrm{res},i}$, and threshold powers, $P_{\mathrm{thr},i}$, are summarised in \cref{table:summary}. We then compared the measured threshold points with the predictions of the model based on~\eqref{eq:eta_nl} and~\eqref{eq:double_resonance}, as shown in \cref{fig:TBD}. For a given set of phase parameters, the model predicts isolated temperatures $T_i$ at which the fundamental and second-harmonic fields are simultaneously resonant. At each model-predicted doubly resonant temperature $T_i$, the corresponding threshold power is modelled as
\begin{equation}
    P_i
    =
    \frac{P_{\mathrm{thr}}^{\mathrm{opt}}}
    {\eta_{\mathrm{nl}}(T_i,\Phi_{\mathrm{L}},\Phi_{\mathrm{R}})} .
    \label{eq:threshold_model}
\end{equation}
Here, $\eta_{\mathrm{nl}}$ is the normalized nonlinear interaction strength defined in \cref{eq:eta_nl}. The parameter $P_{\mathrm{thr}}^{\mathrm{opt}}$ is a fitted threshold scale corresponding to the hypothetical threshold of the same OPO for $\Phi_{\mathrm{R}}=\Phi_{\mathrm{L}}=0$. This parameter is particularly useful because it quantifies how much the measured threshold is increased relative to the ideal case. In this comparison, all other cavity conditions, including optical loss, mode matching, and pump coupling, are assumed to remain unchanged. The phase $\Phi_{\mathrm{R}}$ was kept fixed at the value inferred from the independent double-pass SHG measurement, so that the remaining fit parameters were $\Phi_{\mathrm{L}}$ and $P_{\mathrm{thr}}^{\mathrm{opt}}$.

To quantify the agreement between the measured threshold points and the corresponding model predictions, we defined the cost function
\begin{equation}
\mathcal{C}(\Phi_{\mathrm{L}},P_{\mathrm{thr}}^{\mathrm{opt}})
=
\sum_{i}
\left[
\left(\frac{P_i-P_{\mathrm{thr},i}}{\sigma_P}\right)^2
+
\left(\frac{T_i-T_{\mathrm{res},i}}{\sigma_T}\right)^2
\right] .
\label{eq:fit_OPO}
\end{equation}
Here, $(T_i,P_i)$ denote the corresponding model-predicted doubly resonant solutions for a given parameter set. Since the model predicts isolated resonance solutions rather than a continuous curve, $\mathcal{C}$ was evaluated for all neighbouring model-solution pairs, and the minimum value was retained. The weighting parameters $\sigma_T$ and $\sigma_P$ define the estimated experimental tolerances used to weight deviations in temperature and threshold power. In this work, we used $\sigma_T = \SI{1}{\celsius}$ and $\sigma_P = \SI{3}{\milli\watt}$.

To visualise the propagated fit sensitivity, we considered all parameter sets satisfying
\begin{equation}
\mathcal{C}(\Phi_{\mathrm{L}},P_{\mathrm{thr}}^{\mathrm{opt}})
\leq
\mathcal{C}_{\min}+\Delta\mathcal{C},
\end{equation}
with $\Delta\mathcal{C}=2.30$, chosen by using the numerical value of the joint 68\% $\Delta\chi^2$ criterion for two fit parameters as a convenient reference scale. Since $\sigma_T$ and $\sigma_P$ represent estimated experimental tolerances rather than independently determined statistical uncertainties, the resulting envelope should be interpreted as a propagated fit-sensitivity band rather than as a strict statistical confidence interval. The accepted parameter ranges were $\Phi_{\mathrm{L}}/\pi=0.343$--$0.372$, $1.196$--$1.205$, and $1.628$--$1.634$, and $P_{\mathrm{thr}}^{\mathrm{opt}}=\SIrange{8.5}{13.0}{\milli\watt}$, $\SIrange{22.2}{24.3}{\milli\watt}$, and $\SIrange{15.1}{17.2}{\milli\watt}$ for OPO$_{\mathrm{A}}$, OPO$_{\mathrm{B}}$, and OPO$_{\mathrm{C}}$, respectively. For each accepted parameter set, the model curve $P_{\mathrm{thr}}(T)$ was recalculated, and the shaded envelope was obtained from the minimum and maximum predicted threshold power at each temperature. The width of the envelope depends on the accepted parameter range for each OPO; broader accepted variations in $\Phi_{\mathrm{L}}$ and $P_{\mathrm{thr}}^{\mathrm{opt}}$ lead to a broader propagated model envelope. The envelope is narrow close to the measured threshold points, where the fit is directly constrained by the data, but can become broader in temperature regions away from the measurements, where different accepted parameter sets lead to noticeably different extrapolated threshold values. This effect is further amplified in regions where the threshold curve is steep or where $P_{\mathrm{thr}}(T)\propto 1/\eta_{\mathrm{nl}}(T)$ is particularly sensitive to small changes in the nonlinear-interaction envelope.

The coatings on the PR side of the input coupler and the plano AR side of the crystal introduce additional wavelength-dependent phases of 
$\phi^{\mathrm{PR}}_\text{L}=
\SI{181.2}{\degree}~(\SI{775}{\nano\metre}) 
- \SI{158.5}{\degree}~(\SI{1550}{\nano\metre}) 
= 0.126\pi$ and 
$\phi^{\mathrm{AR}}_\text{L} =
\SI{226.0}{\degree}~(\SI{775}{\nano\metre}) 
- \SI{296.3}{\degree}~(\SI{1550}{\nano\metre}) 
= -0.391\pi$. 
Using the total left-side phase 
$\Phi_\text{L} =
\phi^{\mathrm{PR}}_\text{L} +
2\phi^{\mathrm{AR}}_\text{L}
+\psi_\text{L}$,
together with the phase $\Phi_\text{R}$ extracted from the double-pass SHG measurements, allows us to infer the remaining phase $\psi_\text{L}$ associated with the left part of the crystal-cavity system. The corresponding effective left-edge lengths are \SI{0.20}{\micro\metre}, \SI{10.57}{\micro\metre}, and \SI{3.53}{\micro\metre} for the three systems, respectively.

The three investigated crystal-cavity systems exhibit markedly different behaviours despite their nominally similar designs. For OPO$_\text{A}$, the extracted phases $\Phi_\text{L}$ and $\Phi_\text{R}$ place one of the accessible doubly resonant operating points close to a region of high nonlinear interaction. As a result, although the system does not realise the ideal phase condition, it still operates relatively close to the optimal regime, with a minimum threshold of only $1.52 P_\text{thr}^\text{opt}$. In contrast, OPO$_\text{B}$ exhibits a substantially reduced nonlinear interaction. Interestingly, its inferred right-edge length is closest to the ideal compensation length $\Delta^{\mathrm{ideal}}_\mathrm{r}$, differing by less than \SI{0.5}{\micro\metre}. However, because the achievable gain is determined by the combined phase contributions $\Phi_\text{L}$ and $\Phi_\text{R}$ together with the discrete double-resonance condition, this apparently favourable right-side phase condition does not guarantee optimal OPO operation. In this case, the combined phase contributions shift the accessible doubly resonant temperatures away from the strong gain region, resulting in a minimum reachable threshold of $3.30 P_\text{thr}^\text{opt}$. This places the system close to the theoretical worst-case limit of $1/0.29 \approx \SI{345}{\percent}$ predicted by the model. Our analysis further reveals that the OPO$_\text{C}$ was operated far from its optimal nonlinear interaction regime. While the experimentally accessed operating point at \SI{46.4}{\celsius} corresponds to a threshold of $3.54 P_\text{thr}^\text{opt}$, the model indicates that a lower-threshold doubly resonant point exists near \SI{23}{\celsius}, outside the experimentally used operating range, where the threshold would be reduced to $2.50 P_\text{thr}^\text{opt}$. Our results demonstrate that the phase-sensitive cavity dynamics not only determine the achievable nonlinear efficiency, but also strongly constrain which doubly resonant temperatures provide practical low-threshold operation.

The extracted phase parameters for all three systems are mapped onto the $(\Phi_\mathrm{L},\Phi_\mathrm{R})$ parameter space in \cref{fig:density_plot}, directly linking the experimentally observed threshold behaviour to the global phase-sensitive landscape predicted by the model. Together, these results confirm that nominally similar linear \acp{OPO} can operate in fundamentally different nonlinear regimes solely due to microscopic phase variations originating from crystal edges and reflective coatings.


\section{Discussion and conclusion}

We have investigated phase-sensitive nonlinear interactions in doubly resonant linear standing-wave \acp{OPO} based on periodically poled crystals. By connecting single-pass, double-pass, and cavity-enhanced configurations within a common model, we showed that the effective nonlinear gain is not determined solely by the bulk phase-matching condition. Instead, it is governed by the coherent superposition of forward- and backward-generated fields, making the linear OPO an interference-based nonlinear system whose performance depends on crystal-edge phases, wavelength-dependent coating phases, and the doubly resonant cavity condition.

Using double-pass SHG measurements together with OPO threshold measurements, we extracted the relevant phase contributions in three nominally similar crystal-cavity systems. The measured systems exhibited markedly different threshold behaviours, which can be explained by microscopic phase variations arising from crystal-edge termination and reflective coatings. Micron-scale variations in the effective edge length can translate into $\pi$-scale phase shifts, strongly modifying the temperature dependence of the nonlinear gain and the accessible oscillation thresholds. The model predicts a maximum threshold penalty of approximately $\sim 3.4\times$ for temperature-optimized operation, while the experimentally investigated devices showed threshold variations of up to $\sim 5.9\times$ between nominally similar systems.

These results demonstrate that the optimal operating temperature of a compact linear OPO is not necessarily the bulk phase-matching temperature, but rather one of the discrete doubly resonant temperatures that maximizes the phase-sensitive nonlinear interaction. Phase extraction therefore provides a practical diagnostic tool for identifying favourable low-threshold operating points. At the same time, practical thermal constraints may prevent access to the optimum point in a given device. Moreover, controlling only one cavity-side phase, for example by tuning the temperature near the HR-coated side of the crystal \cite{Schonbeck2018,hagemann202410dBSqueezeLaser}, is generally insufficient, since both crystal-edge phases determine the nonlinear gain while the doubly resonant condition restricts the accessible operating temperatures.

Our findings suggest three complementary routes towards reproducible low-threshold \acp{OPO}. First, phase contributions from coatings and crystal-edge termination should be included at the design and fabrication stage. In particular, coating reflection phases should be specified at both wavelengths, and crystal polishing should control the effective final-domain length whenever possible. Second, residual phase errors after fabrication could be compensated by local temperature tuning near the crystal edges. Two independently controlled edge-temperature regions could provide a lossless actuator for tuning $\Phi_L$ and $\Phi_R$ without introducing additional intracavity optics. Third, the poling period and accessible temperature tuning range should be designed together with the phase-sensitive cavity dynamics in mind. Although a poling period targeting around \SI{35}{\celsius} is safe for an ideal phase-matching bandwidth of approximately \SI{10}{\celsius}, the phase-sensitive response of a linear OPO can shift the practically relevant operating region by up to about \SI{25}{\celsius}. In such cases, the optimum doubly resonant point may even move below room temperature. This suggests that a slightly larger poling period, together with an OPO design allowing a wider temperature tuning range, is desirable for robust low-threshold operation. Such phase-aware design and correction strategies are particularly relevant for scalable photonic architectures, where multiple independent squeezed-light sources must operate with reproducible thresholds and nonlinear gains. More generally, the sensitivity to crystal-edge phases suggests that deliberately specified edge terminations could provide an additional design parameter for engineering the joint spectral amplitude of frequency-nondegenerate OPO output fields.

\section{Acknowledgements}
We acknowledge support from the Danish National Research Foundation (bigQ, DNRF0142), Innovation Fund Denmark (PhotoQ, 1063-00046A and QuantERA-ClusSTAR, 3155-00024A), ERC Advanced grant ClusterQ (no. 101055224), and EU Horizon Europe (CLUSTEC, grant agreement no. 101080173). We thank Roman Schnabel for useful discussions.




\end{document}